\documentclass{elsart}
\usepackage{epsfig}
\usepackage{wrapfig}

\begin{document}

\begin{frontmatter}
\title{Measurements of spin rotation parameter $A$ in pion-proton
elastic scattering at 1.62~GeV/c}

\author{I.G.~Alekseev}, 
\author{P.E.~Budkovsky},
\author{V.P.~Kanavets$^*$}, 
\author{L.I.~Koroleva},
\author{B.V.~Morozov}, 
\author{V.M.~Nesterov}, 
\author{V.V.~Ryltsov},
\author{D.N.~Svirida}, 
\author{A.D.~Sulimov}, 
\author{V.V.~Zhurkin}
\address{Institute for Theoretical and Experimental Physics, \\ 
         B.~Cheremushkinskaya 25, Moscow, 117259, Russia \\
         Tel: 7(095)123-80-72, Fax: 7(095)883-96-01 \\
         E-mail : kanavets@vitep5.itep.ru}
\author{Yu.A.~Beloglazov},  
\author{A.I.~Kovalev},  
\author{S.P.~Kruglov},  
\author{D.V.~Novinsky},
\author{V.A.~Shchedrov},  
\author{V.V.~Sumachev$^*$},  
\author{V.Yu.~Trautman}  
\address{Petersburg Nuclear Physics Institute, \\             
         Gatchina, Leningrad district, 188350, Russia}
\author{N.A.~Bazhanov},
\author{E.I.~Bunyatova}
\address{Joint Institute for Nuclear Research, \\
	 Dubna, Moscow district, 141980, Russia}

{\footnotesize $^*$ Spokenpersons. \hfill}

\begin{abstract}
The ITEP-PNPI collaboration presents the results of the measurements of the 
spin rotation parameter $A$ in the elastic scattering of positive and negative
pions on protons at $P_{\rm beam}=1.62$~GeV/c. 
The setup included a longitudinally-polarized proton target 
with superconductive magnet,
multiwire spark chambers and a carbon polarimeter with thick filter.
Results are compared to the predictions of partial wave analyses.
The experiment was performed at the ITEP proton synchrotron, Moscow.

PACS number(s) : 13.75.Gx, 13.85.Dz, 13.88.+e.

\end{abstract}

\begin{keyword}
elastic pion-nucleon scattering,
spin rotation parameters,
partial wave analysis,
baryon spectroscopy,
polarized target,
carbon polarimeter
\end{keyword}

\date{}

\end{frontmatter}

\section{Introduction}
Both the experimental and the theoretical baryon
spectroscopy is under steady progress in recent years.
The study of photo- and 
electro-production of nonstrange resonances was started
by  ELSA (Bonn) and CLAS (CEBAF) experiments.
As a result of the development of the chiral perturbation theory
(CHPT) dynamically generated resonances in the isospin channels
0($\Lambda(1405)$) and $\frac{1}{2}$($S_{11}(1535)$) were predicted
as a consequence of the existence of quasi-bound $\bar{K} N$ and
$K\Sigma$ states~\cite{Kaizer}. Method of amplitude
speed plots was developed \cite{Holler} for resonance parameter studies.
Nevertheless a number of important questions of the baryon spectroscopy
is still waiting for their solutions. Among them is the existence of 
clusters of baryon resonances, low energy of two-phonon excitation 
(for instance $P_{11}(1440)$), the problem of ``missing'' resonances 
in the second resonance region, the role played by the gluonic degrees of
freedom at low energies.

Partial wave analysis (PWA) is one of the most powerful instruments 
in the baryon spectroscopy. But it is still not free 
from ambiguities caused by the incomplete experimental database. 
Especially poor is the knowledge on spin rotation parameters $A$ and $R$ in 
the region of incident particle momenta above 0.75~GeV/c. Predictions 
of these parameters made by different PWA contradict with each other in a
number of kinematic regions. The single measurement in this region 
fulfilled by the ITEP-PNPI collaboration \cite{RA} contradict with
predictions of {\bf KH80}\footnote{Here and later we use the notations
of PWA given by their authors} \cite{KH} and {\bf CMB} \cite{CMB} and 
agree with that of {\bf SM90} \cite{Arndt}. The analysis of 
the data by the method of transverse amplitude zeroes \cite{Barrelet}
show that the
disagreement with {\bf KH80} and {\bf CMB} could be attributed to
the discrete ambiguity of Barrelet-type.
Absence of the data on spin rotation parameters at the 
time when these analyses were performed didn't allow their authors to choose
the correct branch of the solution. It was shown that proper
amplitude correction results in a good agreement between the predictions
of {\bf KH80} and {\bf CMB} and the experimental data,
but the parameters of 6 resonances with isospin 
$\frac{3}{2}$ and masses near 1.9~GeV/c$^2$ are significantly changed
\cite{SvirThes}.

This experiment was aimed at several goals: (I) to obtain new
experimental data for unambiguous reconstruction of 
$\pi^+p$-elastic amplitudes in the range of 
$\theta_{cm}=120$--$140^o$, where the largest disagreement
between predictions of the existing PWA's is observed;
(II) to confirm the choice of the transverse amplitude
zero trajectory (solution branch) done in our previous analysis \cite{SvirThes};
(III) to test PWA predictions on spin rotation 
parameters in $\pi^-p$-elastic scattering.

\section{Experimental conditions}
Spin rotation parameter $A$ is measured in the elastic scattering on 
the lon\-gi\-tu\-di\-nal\-ly-polarized proton target as a component of
the recoiled proton polarization perpendicular to its momentum and
laying in the scattering plane. This component is determined by the
secondary scattering of the recoiled protons on the carbon 
filter.

\begin{figure}[bt]
\epsfig{figure=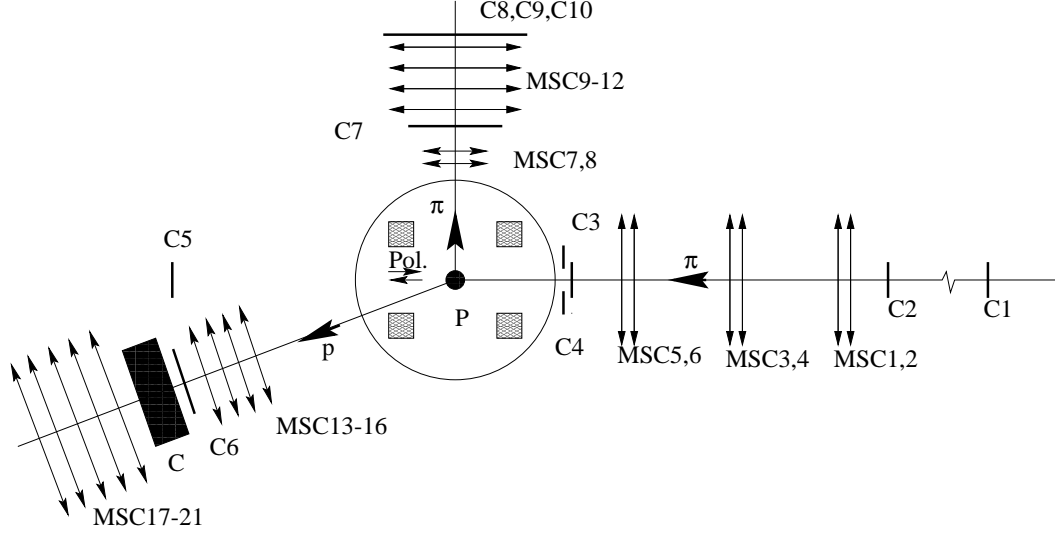,width=\textwidth}
\caption{The experimental layout (not to scale).}
\label{fig:setup}
\end{figure}

The apparatus is shown in Fig.~\ref{fig:setup}.
Its basic elements 
are: (i) polarized proton target (P) \cite{Target}; (ii) carbon 
filter (C); (iii) four sets of multiwire magnetostrictive spark chambers 
to detect the incident beam (MSC1--MSC6), the scattered pion (MSC7--MSC12) and
the recoiled proton before (MSC13--MSC16) and after (MSC17--MSC21) the 
second scattering and (iv) a number of scintillation counters (C1-C10) to 
provide the trigger and to identify the positive pions in the beam by the time of 
flight. A container filled with the target material (propanediole
$\mathrm{C_3H_8O_2}$ doped by $\mathrm{Cr^V}$ complexes) is placed
into magnetic field of 2.5~T created by a Helmholtz pair of 
superconductive coils. The container has a cylindrical form with
vertical size and diameter of 30~mm~$\times$~30~mm.
Cooling of the target downto
0.5~K is provided by an evaporation-type $^3$He cryostat.
The polarization is pumped by the dynamic nuclear
orientation method up to the absolute value of 70-80~\% with the measurement
uncertainty 2~\%. 

The two sets of chambers with the 36.5~g/cm$^2$ thick carbon filter 
in between form the polarimeter. The analyzing power of this 
polarimeter was measured in advance at the polarized proton 
beam of the ITEP accelerator \cite{Polarim}. The false asymmetry
in the polarimeter was also measured with a pion beam and appeared to be
$0.0026\pm0.0014$.

This paper presents the results of two runs at the ITEP accelerator.
During the first run $1.4 \times 10^6$ triggers with 
$\pi^+$ beam were obtained while the second one gave 
$6.5 \times 10^5$ triggers from $\pi^-$ beam.

\section{Data processing}

\begin{wrapfigure}{r}{0.5\textwidth}
\epsfig{figure=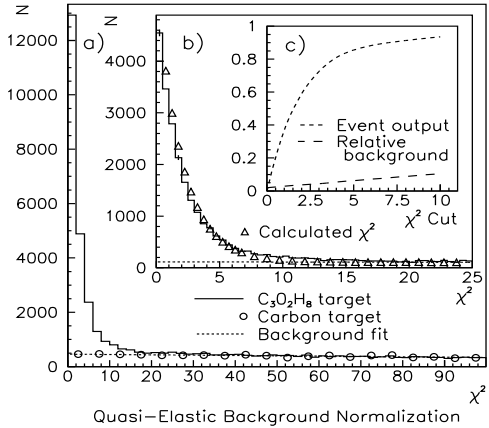,width=0.47\textwidth,height=0.42\textwidth}
\caption{(a) $\chi^2$ distribution of the events on the polarized target 
(solid lines) and on the carbon target (open dots).
(b) $\chi^2$ distribution of $\pi p$ elastic events (solid lines)
and expected $\chi^2$ distribution (open triangles).
(c) Event output and relative background vs $\chi^2$ cut value.}
\label{fig:chi2}
\end{wrapfigure}

The processing of the data was performed in several steps:
\begin{itemize}
\item Events of the elastic $\pi p$ scattering on the polarized
target were selected by coplanarity and pion-proton
angular correlation. 
$\chi^2=(\frac{\Delta \phi}{\sigma_\phi})^2 + 
(\frac{\Delta \theta}{\sigma_\theta})^2$ was calculated for
each event, where $\Delta \phi$ and $\Delta \theta$ are 
deviations from the elastic kinematics in azimuthal and polar
angles and $\sigma_\phi$ and $\sigma_\theta$ are the RMS of
the corresponding distributions obtained from Monte Carlo
simulation. $\chi^2$ distributions of the data taken
on the polarized target and on a carbon target normalized
at $\chi^2 > 35$ are shown in Fig.~\ref{fig:chi2}a.
The
distribution from the carbon target representing the pure quasielastic
background can be approximated with a straight line.
The distribution of elastic events with subtracted background
follows $\chi^2$ distribution with 2 degrees of freedom 
(Fig.~\ref{fig:chi2}b). The selection criteria $\chi^2 < 5$
corresponds to 6\% background and 85\% good events after the cut.
The normal polarization of the quasielastic background
was taken 0.7 of that for the elastic scattering \cite{Polarim}.
For every event a $3 \times 3$ matrix was calculated, describing recoiled
proton spin rotation in the magnetic field of the setup along its trajectory
from the vertex of the first scattering to the point of the rescattering
on the carbon nucleus.
\item Single track events with polar angle of the second
scattering $> 3^o$ were selected in the polarimeter.
From them only those events were taken for which all the azimuthal
angles are allowed by the chambers geometry. The average
analyzing power for selected events is 0.191. More details
on data processing in the polarimeter can be found in
\cite{Polarim}. After this selection 16686 events of
elastic $\pi^+p$ scattering and 4708 events of $\pi^-p$
scattering were left for further processing.
\item The method of maximum likelihood was used to get
the polarization parameters from the data. The probability density
was built only as a function of parameters $A$ and $P$, while the parameter 
$R$ was calculated from the equation: 
$P^2 + A^2 + R^2 = 1$. The result of the fit practically
does not depend on the assumption over the sign of $R$.
The likelihood function accounts for: the target polarization,
the quasielastic background and its polarization, the analyzing power
of the polarimeter and the rotation of the proton spin
in the magnetic field between the first and the second 
scattering.
\end{itemize}

\section{Results}

\begin{table}[h]
\caption{Polarization and spin rotation parameters in $\pi p$ elastic 
scattering at 1.62~GeV/c.}
\begin{center}
\begin{tabular}{|c|c|c|c|c|}
\hline
\multicolumn{2}{|c|}{$\theta_{cm}$ (degr.)} & $P$ & $A$ & $|R|$ \\
\cline{1-2}
range & mean &&&\\
\hline
\multicolumn{5}{|c|}{$\pi^+p$ elastic scattering}\\
\hline
118--123.5 & 121.7 & $0.24\pm 0.12$ & $0.27\pm 0.18$  & $0.93\pm 0.06$\\
123.5--127 & 125.2 & $0.30\pm 0.12$ & $0.36\pm 0.20$  & $0.88\pm 0.09$\\
127--131   & 128.8 & $0.40\pm 0.13$ & $-0.32\pm 0.20$ & $0.86\pm 0.10$\\
131--140   & 133.6 & $0.29\pm 0.13$ & $-0.40\pm 0.21$ & $0.87\pm 0.11$\\
\hline
\multicolumn{5}{|c|}{$\pi^-p$ elastic scattering}\\
\hline
118--124.8   & 122.3 & $-0.11\pm 0.19$ & $0.88\pm 0.28$ & $0.46\pm 0.54$\\
124.8--129.4 & 127.0 & $0.03\pm 0.19$  & $0.56\pm 0.28$ & $0.83\pm 0.19$\\
129.4--140   & 132.8 & $0.19\pm 0.20$  & $0.51\pm 0.29$ & $0.84\pm 0.18$\\
\hline
\end{tabular}
\end{center}
\label{tab:PRA}
\end{table}

\begin{figure}[th]
\begin{tabular}{cc}
\epsfig{figure=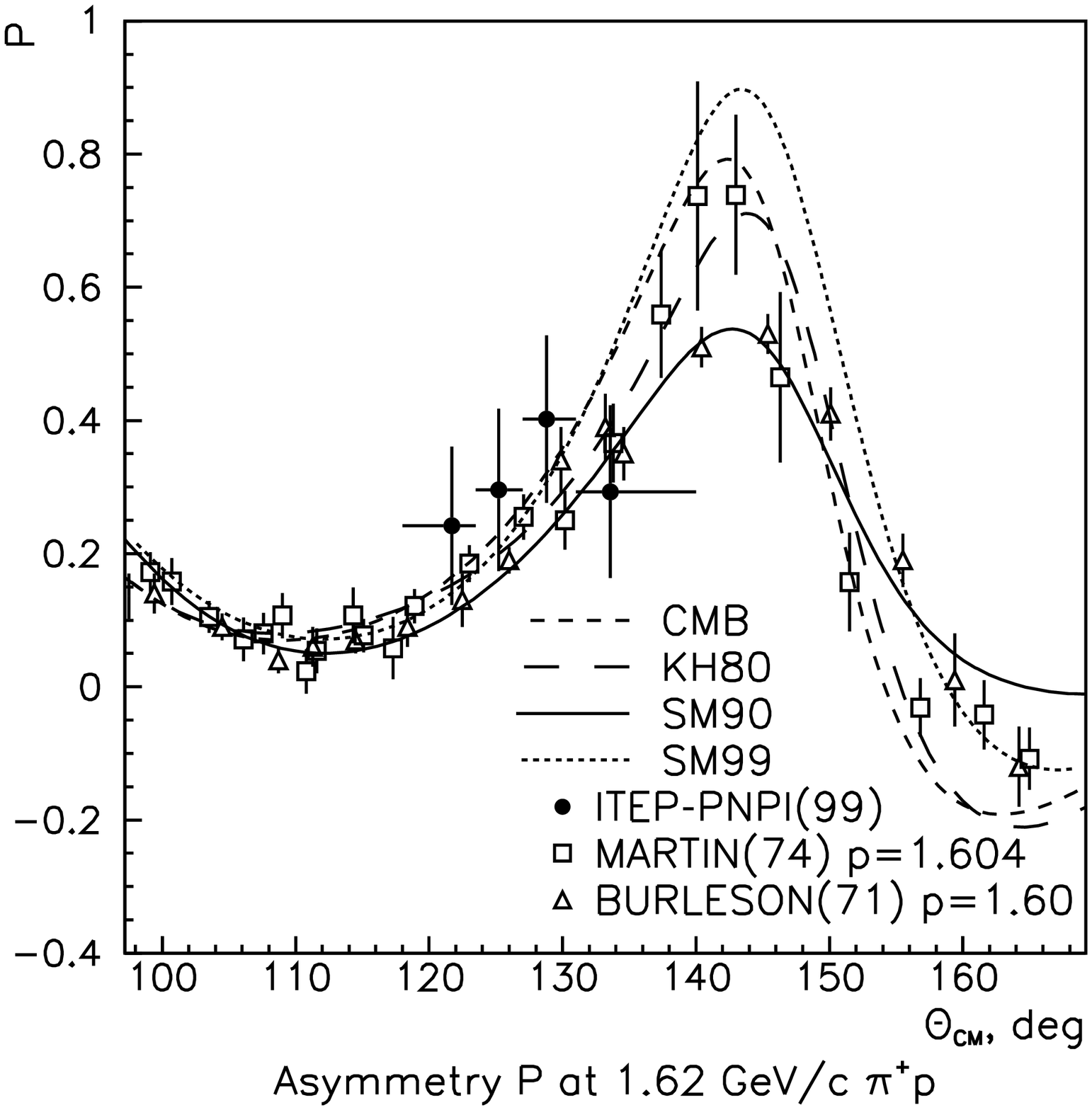,width=0.47\textwidth,height=0.42\textwidth}&
\epsfig{figure=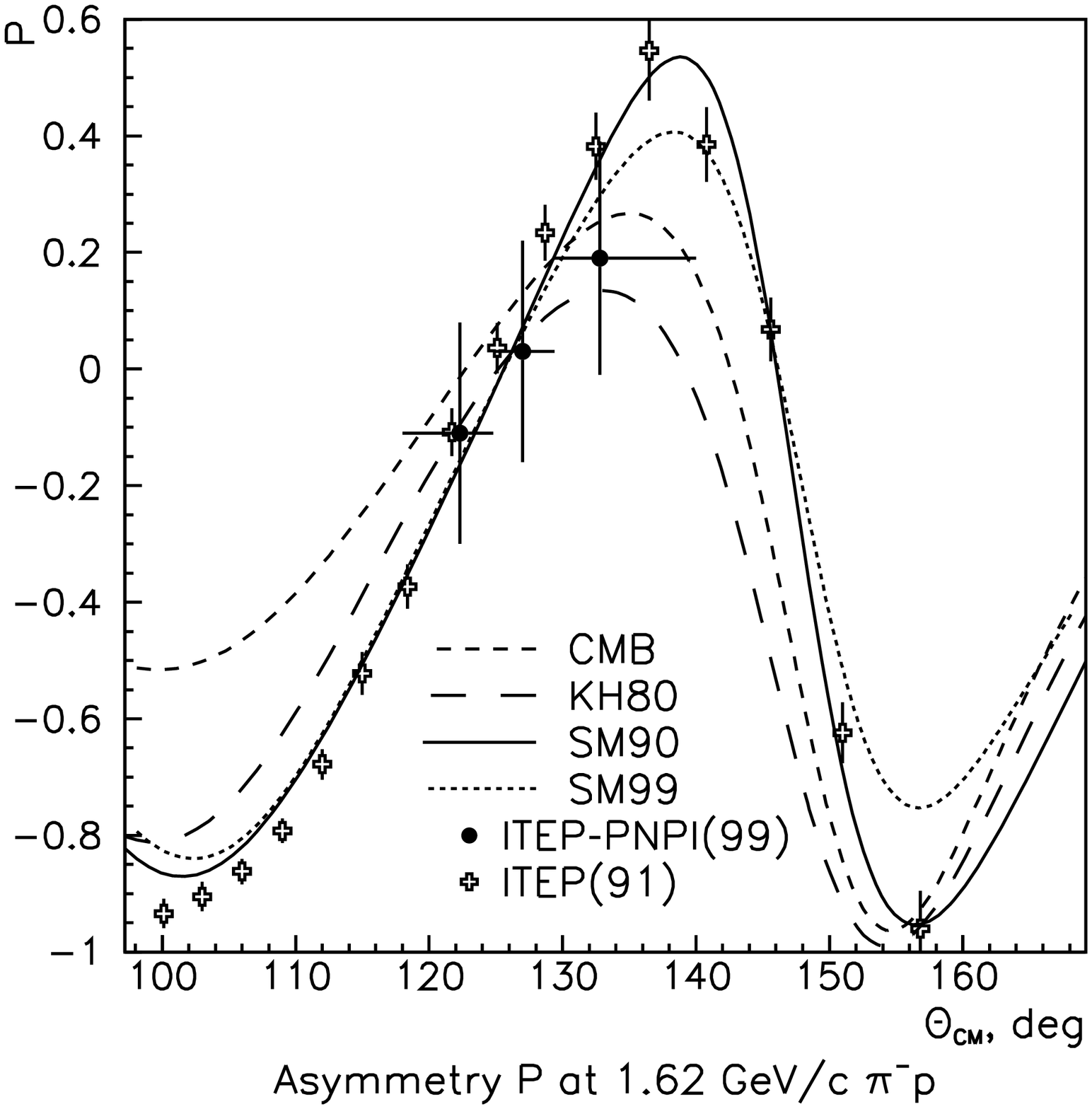,width=0.47\textwidth,height=0.42\textwidth}\\
a) & b) \\
\epsfig{figure=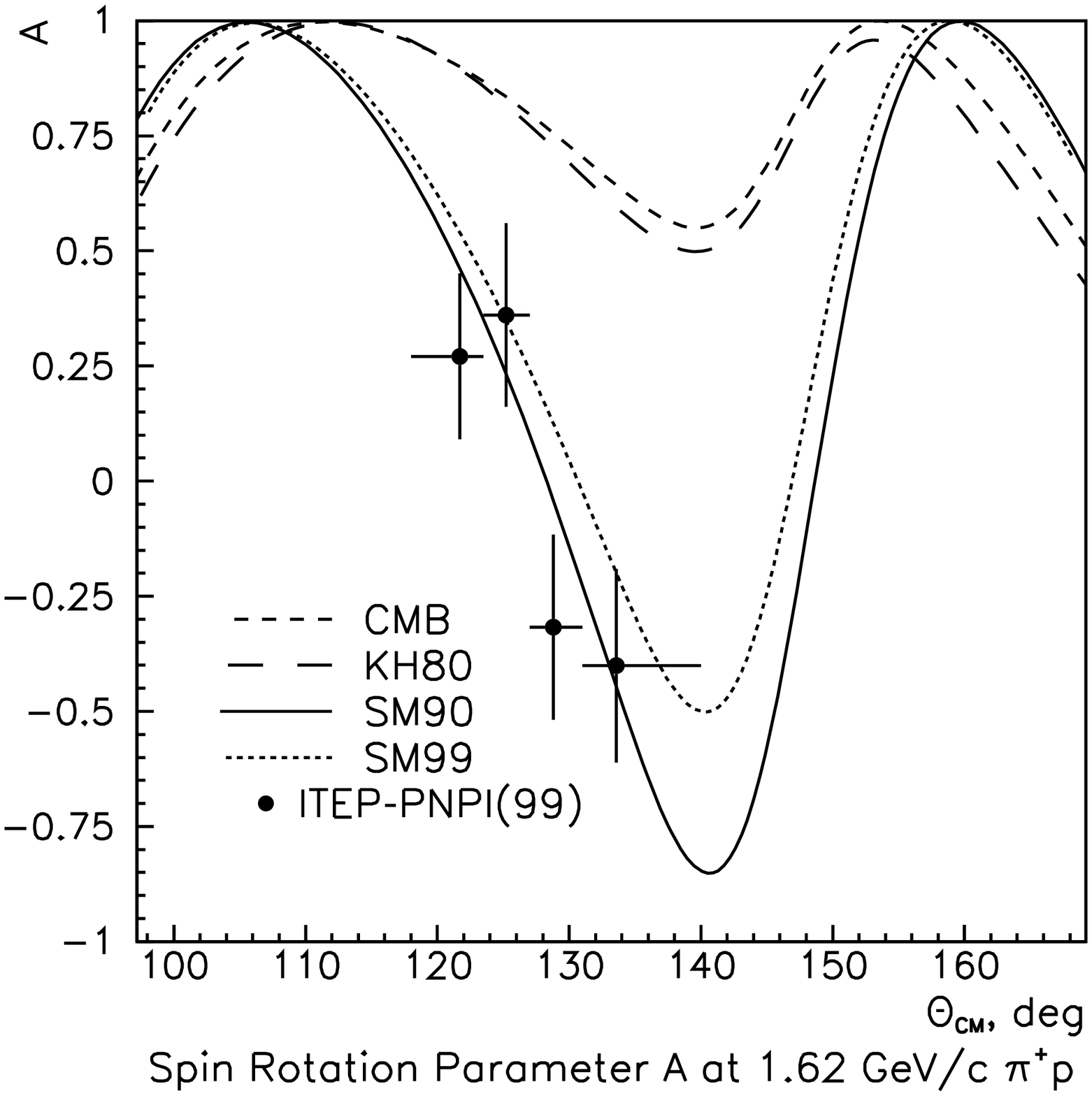,width=0.47\textwidth,height=0.42\textwidth}&
\epsfig{figure=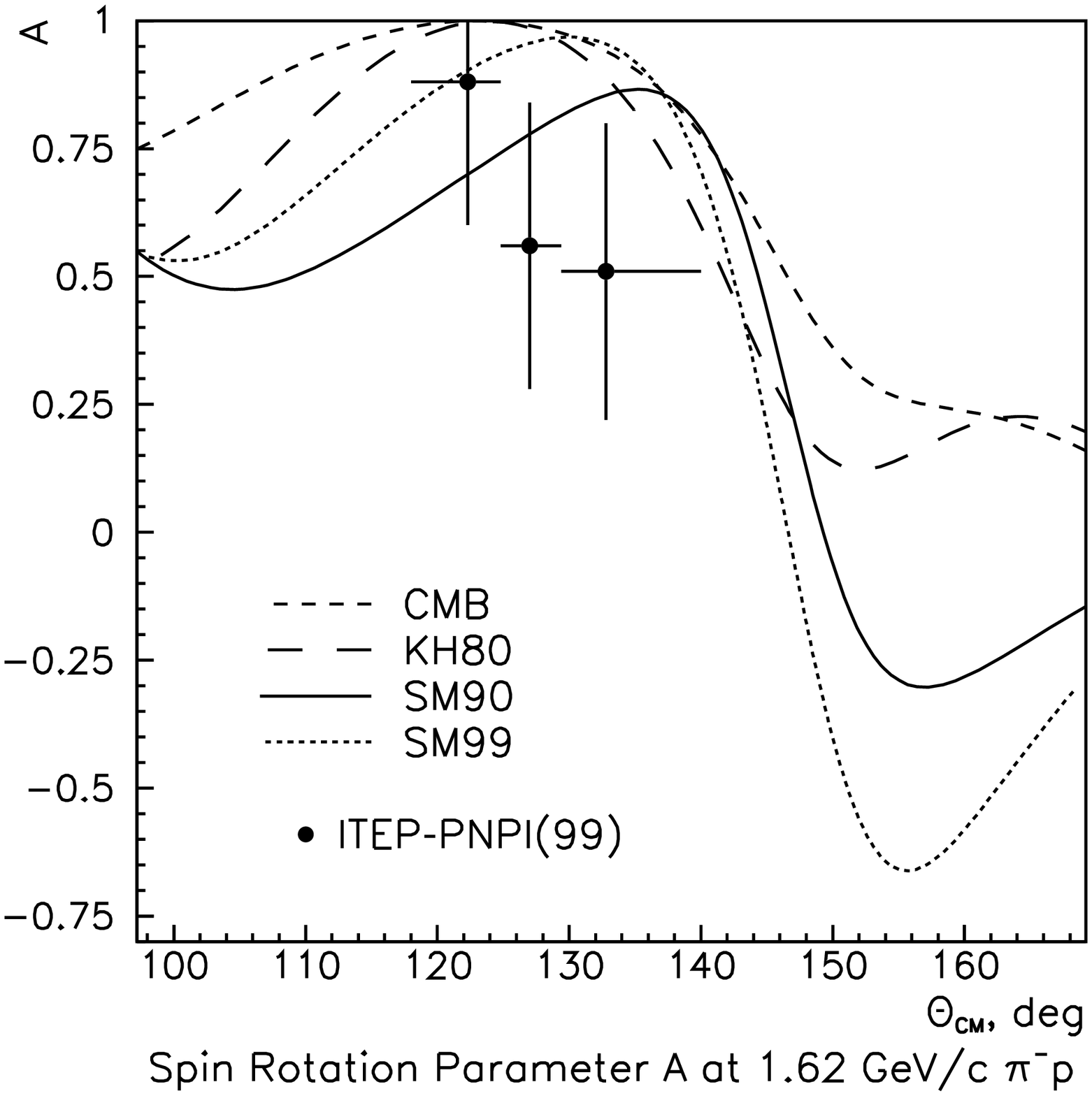,width=0.47\textwidth,height=0.42\textwidth}\\
c) & d) \\
\epsfig{figure=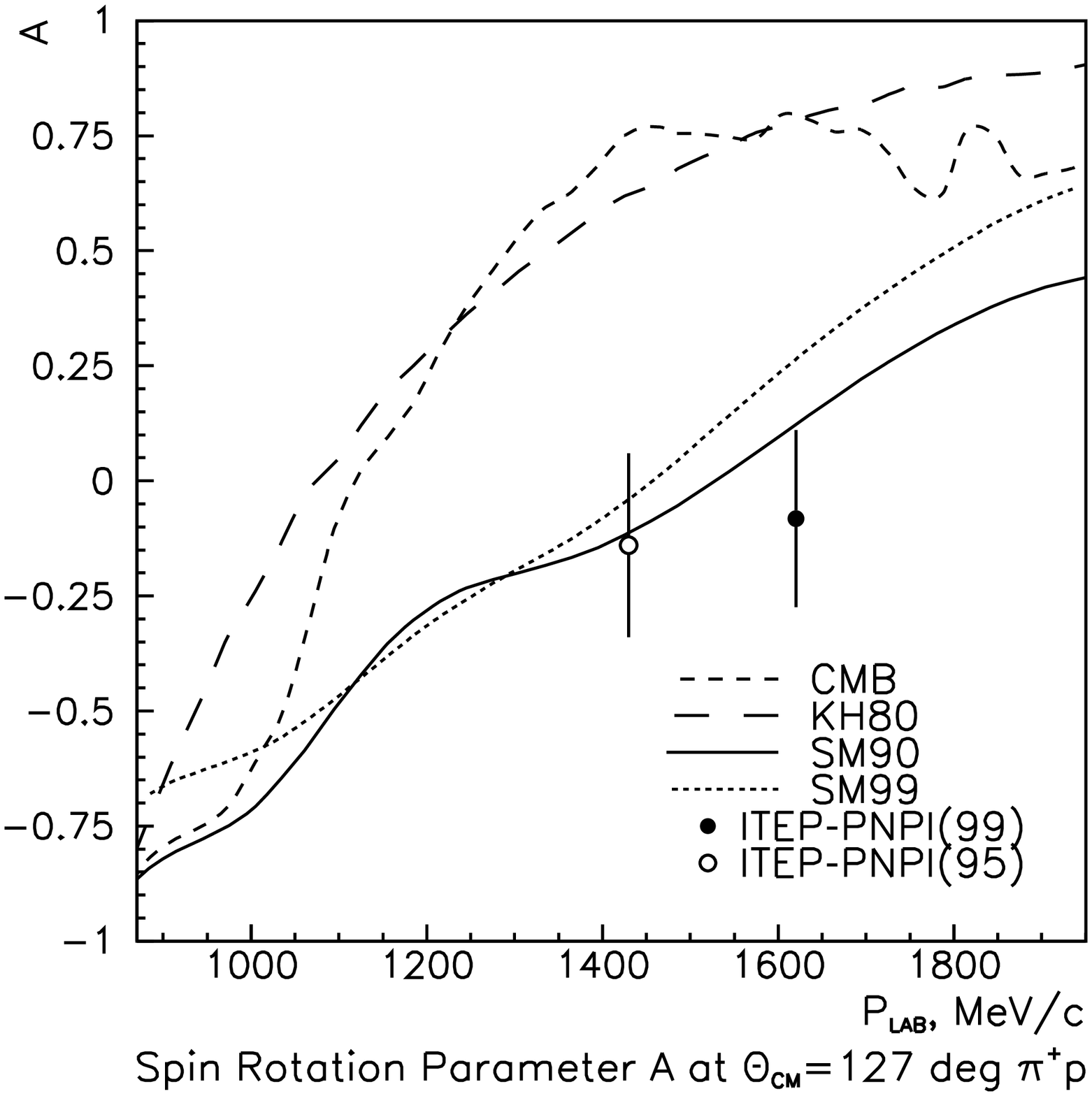,width=0.47\textwidth,height=0.42\textwidth}&
\epsfig{figure=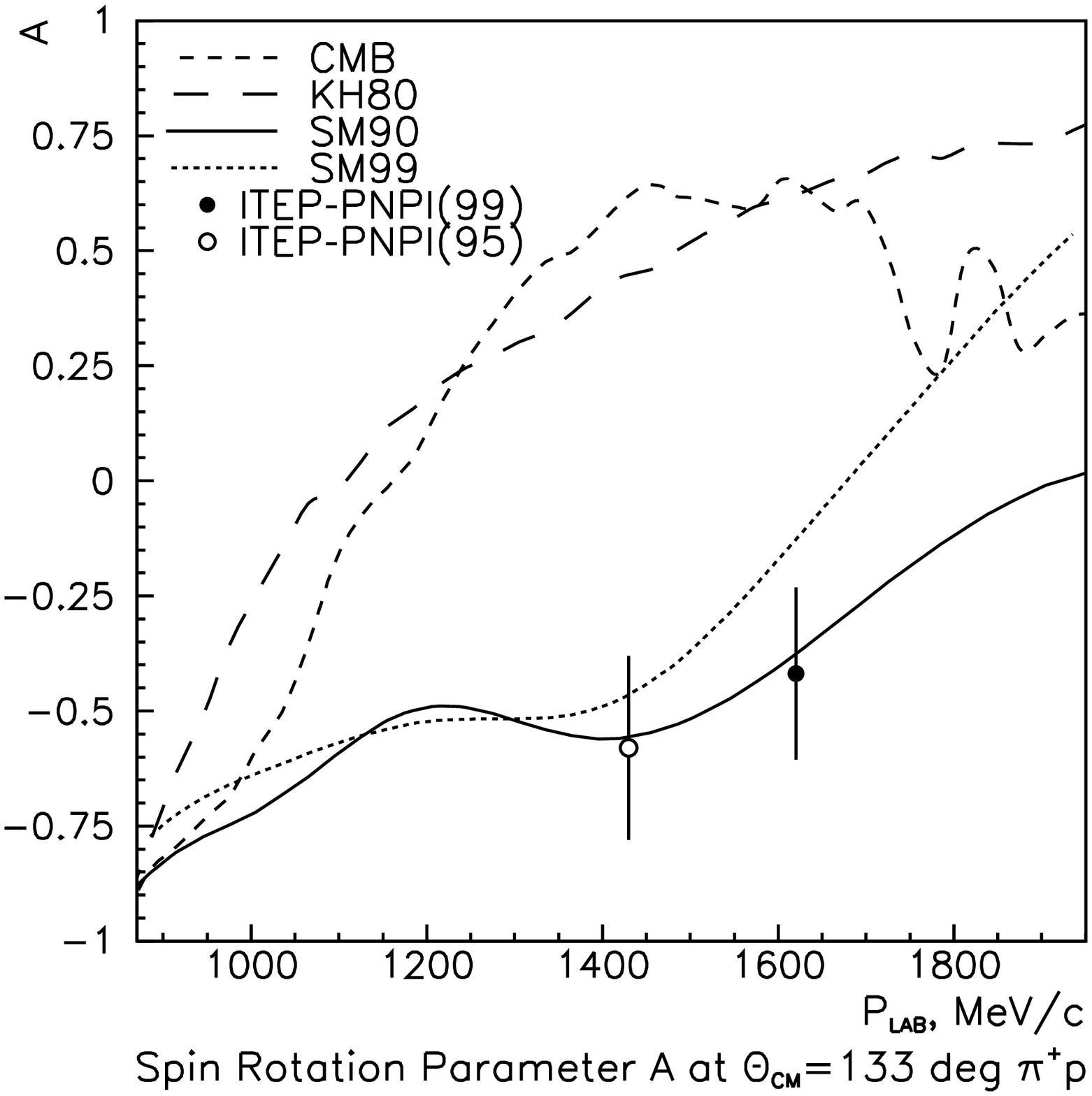,width=0.47\textwidth,height=0.42\textwidth}\\
e) & f)
\end{tabular}
\caption{Results of this work (full dots) compared to the
data from \cite{RA} (open dots), \cite{ITEPPol} 
(open crosses), \cite{Martin} (open squares),
\cite{Youkosava} (open triangles) and predictions
of selected PWA \cite{KH,CMB,Arndt}.
Polarization $P$ at 1.62~GeV/c in $\pi^+p$ (a)
and $\pi^-p$ (b) elastic scattering.
Spin rotation parameter $A$  at 1.62~GeV/c in 
$\pi^+p$ (c) and $\pi^-p$ (d) elastic scattering.
Spin rotation parameter $A$ at $\theta_{cm}=127^o$
(e) and $133^o$ (f) in $\pi^+p$ elastic scattering.}
\label{fig:PRA}
\end{figure}

The results of the experiment are given in the Table~\ref{tab:PRA}.
Only statistical errors are given. All the systematic errors
such as false setup asymmetry, uncertainties in the target 
polarization, analyzing power, amount and polarization of
the background are negligible compared to the statistical
errors. The results on the normal polarization $P$ do not contradict 
within the errors to the results of other works 
\cite{Martin,Youkosava,ITEPPol} and predictions of PWA's 
\cite{KH,CMB,Arndt} (Fig.~\ref{fig:PRA}a,b). 
New results for the
parameter $A$ are shown in Fig.~\ref{fig:PRA}c,d. 
Our data for $\pi^+p$ scattering does not contradict to the 
predictions given by the analyses {\bf SM90} and {\bf SM99} 
\cite{Arndt} and is in strong disagreement with the predictions of
{\bf KH} \cite{KH} and {\bf CMB} \cite{CMB}. This remains true
in a wide momentum range as seen from Fig.~\ref{fig:PRA}e,f, 
where the results of this work are shown together with the data at 
$P_\mathrm{beam}=1.43$~GeV/c from our previous work 
\cite{RA} for two angles $\theta_{cm}=127$ and $133^o$. 
Thus we confirm the conclusion of Ref.~\cite{SvirThes} that
the difference between several PWA comes from Barrelet 
ambiguity and from the choice of right trajectory of
transverse amplitude zeroes. In $\pi^-p$ scattering
the parameter $A$ from this experiment does not
deviate much from PWA predictions, but looks to
be more close to {\bf SM90} and {\bf SM99}.

This work was partially supported by the Russian
Fund for Basic Research grant 99-02-16635 and Russian
State Scientific Technical Program "Fundamental Nuclear
Physics".

\end{document}